\documentclass[twocolumn,showpacs,preprintnumbers,amsmath,amssymb]{revtex4}
\usepackage{graphicx}
\usepackage{dcolumn}
\usepackage{bm}
\begin{document}

\newcommand{\RE}{\mbox{Re}}
\newcommand{\etal}{{\it et~al}}
\newcommand{\IM}{\mbox{Im}}

\title{Theory of high harmonic generation in relativistic laser
       interaction with overdense plasma}

\author{T.~Baeva$^{1}$, S.~Gordienko$^{1,2}$, A.~Pukhov$^1$}

\affiliation{$^1$Institut f\"ur Theoretische Physik I,
            Heinrich-Heine-Universit{\"a}t D\"usseldorf, D-40225, Germany \\
            $^2$L.~D.~Landau Institute for Theoretical Physics, Moscow, Russia
           }

\date{\today}

\begin{abstract} {
 \noindent
High harmonic generation due to the interaction of a short
ultra relativistic laser pulse with overdense plasma is studied
analytically and numerically. On the basis of the ultra relativistic
similarity theory we show that the high harmonic spectrum is
universal, i.e. it does not depend on the interaction details.
The spectrum includes the power law part $I_n\propto n^{-8/3}$ for
$n<\sqrt{8\alpha}\gamma_{\max}^3$, followed by exponential
decay. Here $\gamma_{\max}$ is the largest relativistic $\gamma$-factor
of the plasma surface and $\alpha$ is the second derivative of
the surface velocity at this moment. The high harmonic cutoff at
 $\propto \gamma_{\max}^3$ is parametrically larger than the $4
 \gamma_{\max}^2$ predicted by the ``oscillating mirror'' model based
on the Doppler effect. The cornerstone of our theory is the new physical
phenomenon: spikes in the relativistic $\gamma$-factor of the
plasma surface. These spikes define the high harmonic spectrum
and lead to attosecond pulses in the reflected radiation.
}
\end{abstract}


\maketitle


\section{Introduction}
\label{Intro}



High harmonic generation (HHG) from relativistically intense laser
pulses interacting with solid targets has been identified as a
promising way to generate bright ultra short bursts of X-rays
\cite{Gordienko2004,NJP06,Watts2002}. For the first time this
spectacular phenomenon was observed with nanosecond pulses of long
wavelength ($10.6~\mu$m) CO$_2$ laser light \cite{Carman1981}.

Short after the experimental observation in 1981,
Bezzerides~\etal~studied the problem of harmonic light emission
theoretically 
\cite{Bezzerides1982}.  Their approach based on non-relativistic
equations of motion and hydrodynamic approximation for the plasma
predicted a cutoff of the harmonic spectrum at the plasma
frequency.

10 years later, in 1993, a new approach to the interaction of an
ultra short, relativistically strong laser pulse with overdense
plasma was proposed by Bulanov \etal~\cite{Bulanov1993}. They
"interpreted the harmonic generation as due to the Doppler effect
produced by a reflecting charge sheet, formed in a narrow region at
the plasma boundary, oscillating under the action of the laser
pulse"~\cite{Bulanov1993}. The "oscillating mirror" model predicts a
cutoff harmonic number of $4\gamma^2_{\max}$, where $\gamma_{\max}$
is the maximal $\gamma$-factor of the mirror.

At the beginning of 1996 numerical results of particle-in-cell
simulations of the harmonic generation by femtosecond laser-solid
interaction were presented by P.~Gibbon \cite{Gibbon1996}. He
demonstrated numerically that the high harmonic spectrum goes well
beyond the cutoff predicted in \cite{Bezzerides1982} and also
presented a numerical fit for the spectrum, which approximated the
intensity of the $n$-th harmonic as $I_n\propto n^{-5}$. At about
the same time the laser-overdense plasma interaction was also
studied by Lichters \etal~\cite{Lichters1996}.

The same year the analytical work by von der Linde and Rzazewski
\cite{vonDerLindeRzazewski1996} appeared. The authors used the
"oscillating mirror" model and approximated the oscillatory motion
of the mirror as a $\sin$-function of time without analysis of
the applicability of this approximation. With the explicit form of
the mirror motion an analytical formula for the harmonic spectrum
was obtained.

The first try to describe analytically the high harmonics generated
at the boundary of overdense plasma by a short ultra intense laser
pulse in a universal way that does not rely on an explicit formula
for the exact plasma mirror motion was made in \cite{Gordienko2004}.
This work proposed for the first time the idea of universality of
the harmonic spectrum. In an attempt to advocate the "oscillating
mirror" model and the role of the relativistic Doppler effect the
authors of \cite{Gordienko2004} made use of the steepest descent method
and estimated the harmonic spectrum as $I_n\propto n^{-5/2}$. However
they did not discuss the area of applicability of the steepest descent
method for the specific problem. We are showing in this article that 
applying the straightforward steepest descent method is equivalent to
applying the "oscillating mirror" model that proves to be insufficient
for the treatment of laser-overdense plasma interaction.  In the present
work we revise \cite{Gordienko2004} and clarify the physical picture of HHG.


For the last couple of years the interest in the process of high
harmonic generation from plasmas has enjoyed a revival thanks to the
increasing interest in attoscience.  The recent impressive progress 
in the physics of attosecond X-ray pulses \cite{AttoKrausz} triggers
the fascinating question whether a range of even shorter ones is
achievable with the contemporary or just coming experimental
technology.

L.~Plaja \etal~\cite{Plaja1998} were the first to realize that the
simple concept of the "oscillating mirror" gives an opportunity to
produce extremely short pulses and presented a numerical proof that
the radiation generated by oscillating plasma surfaces comes in the
form of subfemtosecond pulses. For the first time the idea to use
the plasma harmonics for the generation of subattosecond pulses
(zeptosecond range) was announced in \cite{Gordienko2004}.


The present article improves the analytical results of
\cite{Gordienko2004} and studies in detail the ultra relativistic
plasma surface motion. We show that the high harmonics are generated
due to sharp spikes in the relativistic $\gamma-$factor of the
plasma surface. This new physical phenomenon leads to the spectrum
cutoff at the harmonic number $n_\text{cutoff}$

\begin{equation}
n_\text{cutoff} = \sqrt{8\alpha}\gamma_{\max}^3. \label{ap:cutoff}
\end{equation}

\noindent
The cutoff (\ref{ap:cutoff}) is much higher than the
$4\gamma_{\max}^2$-cutoff predicted by the simple 
"oscillating mirror" model. Here $\gamma_{\max}$ is the maximum
$\gamma-$factor of the surface and $\alpha$ is a numerical factor of
the order of unity, related to the plasma surface acceleration.

Our analysis significantly exploits the relativistic plasma
similarity theory \cite{Bubble}, which was developed after the work
\cite{Gordienko2004} had been published. The similarity theory
enables us to rectify our previous results and to present them in a
straightforward and clear way.

In this article we first discuss the physical picture of high
harmonic generation. The production of high harmonics is attributed
to the new physical effect of the relativistic spikes. Then, 
we develop analytical theory describing the spectrum of the high
harmonics and show that this spectrum is universal with slow
power law decay. Finally, the theory is confirmed by direct
particle-in-cell (PIC) simulation results.


\section{Physical Picture of HHG at Overdense Plasma Boundary}
\label{PhysPict}


In this Section we state the problem of high harmonic generation
at the boundary of overdense plasma and qualitatively describe its
main features, which will find their analytical and numerical
confirmation in what follows.

\begin{figure}
  \centerline{\includegraphics[width=8cm,clip]{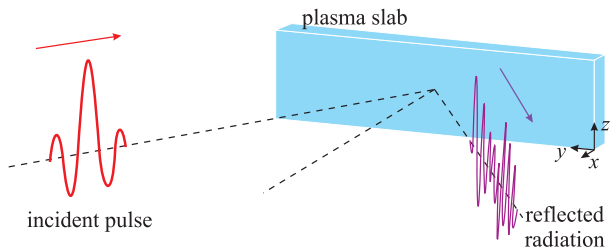}}
  \caption{Geometry of the problem. The laser pulse is
           moving towards the overdense plasma slab, $x$ is perpendicular to the
           surface, $y$ and $z$ are parallel to it.} 
 \label{GeometryProblem}
\end{figure}

Let us consider a short laser pulse of ultra relativistic intensity,
interacting with the sharp surface of an overdense plasma slab (see
Fig.~\ref{GeometryProblem}).

\begin{figure}
  \centerline{\includegraphics[width=4cm,clip]{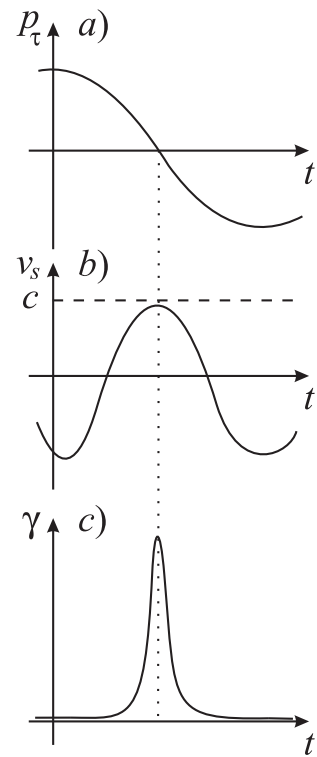}}
 \caption{a) Electron momentum component parallel to
           the surface as a function of time b) The velocity of the plasma
           surface $v_s$ is a smooth function of time, unlike the
           $\gamma$-factor of the surface c)}
 \label{VelSurf}
\end{figure}

We assume that the incident laser pulse is short, so that we can neglect
the slow ion dynamics and consider the electron motion only. The
electrons are driven by the laser light pressure, a restoring
electrostatic force comes from the ions. As a consequence, the
plasma surface oscillates and the electrons gain a normal momentum
component.

Since the plasma is overdense, the incident electromagnetic wave is
not able to penetrate it. This means that there is an electric
current along the plasma surface. For this reason, the momenta of
electrons in the skin layer have, apart from the components
normal to the plasma surface, also tangential components.

According to the relativistic similarity theory \cite{Bubble}, both
the normal and tangential components are of the order of
the dimensionless electromagnetic potential $a_0$. Consequently,
the actual electron momenta make a finite angle with the plasma
surface for most of the times.

Since we consider a laser pulse of ultra relativistic intensity, the
motion of the electrons is ultra relativistic. In other words, their
velocities are approximately $c$. Though the motion of the plasma
surface is qualitatively different: its velocity $v_s$ is not
ultra relativistic for most of the times but smoothly approaches $c$
only when the tangential electron momentum vanishes (see
Fig.~\ref{VelSurf}b).

The $\gamma$-factor of the surface $\gamma_s$ also shows specific
behavior. It has sharp peaks at those times for which the velocity
of the surface approaches $c$ (see Fig.~\ref{VelSurf}c). Thus,
while the velocity function $v_s$ is characterized by its
smoothness, the distinctive features of $\gamma_s$ are its
quasi-singularities.

When $v_s$ reaches its maximum and $\gamma_s$ has a sharp peak, high
harmonics of the incident wave are generated and can be seen in the
reflected radiation. Physically this means that the high harmonics
are due to the collective motion of bunches of fast electrons moving
towards the laser pulse \cite{TPBDiplomaThesis}.

These harmonics have two very important properties. First, their
spectrum is universal. The exact motion of the plasma surface can be
very complicated, since it is affected by the shape of the laser
pulse and can differ for different plasmas. Yet the qualitative
behavior of $v_s$ and $\gamma_s$ is universal, and since it governs
the HHG, the spectrum of the high harmonics does not depend on the
particular surface motion.

\begin{figure} [h]
  \centerline{\includegraphics[width=7cm,clip]{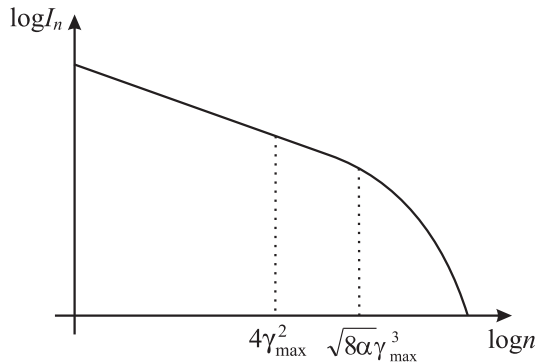}}
 \caption{The universal high harmonic spectrum
           contains power law decay and exponential decay (plotted in log-log
           scale).} 
 \label{SpectrumPlasma}
\end{figure}

We show below that the high harmonic spectrum contains two
qualitatively different parts: power law decay and exponential
decay (see Fig.~\ref{SpectrumPlasma}). In the power law part the
spectrum decays as

\begin{equation}
 I_n\propto 1/n^{8/3},
\end{equation}

\noindent
up to a critical harmonics number that scales as
$\gamma_{\max}^3$, where $I_n$ is the intensity of the $n$th
harmonic (see Section \ref{SectionUniversalSpectrum}). Here
$\gamma_{\max}$ is the maximal $\gamma$-factor of the point, where
the component of the electric field tangential to the surface
vanishes (see Section \ref{BoundCond}).

The second important feature of the high harmonics is that they are
phase locked. This observation is of particular value, since it
allows for the generation of attosecond and even subattosecond
pulses \cite{Gordienko2004}.


\section{Ultra relativistic similarity and the plasma surface motion}
\label{KinDescr}


The analytical theory presented in this work is based on the
similarity theory developed in \cite{Bubble} for collisionless
ultra relativistic laser-plasma regime and is valid both for under- and
overdense plasmas. 

The ultra relativistic similarity theory states that when the
dimensionless laser vector potential ${\bf a}_0=e {\bf A}_0/mc^2$ is
large ($a_0^2 \gg 1$) the plasma electron dynamics
does not depend on $a_0$ and the plasma electron density $N_e$
separately. Instead they merge in the single dimensionless
similarity parameter $S$ defined by

\begin{equation} \label{Sparameter}
 S=\frac{N_e}{a_0 N_c},
\end{equation}

\noindent
where $N_c =\omega_0^2 m/4\pi e^2$ is the critical electron density
for the incident laser pulse with amplitude $a_0$ and carrier frequency
$\omega_0$.

In other words, when the plasma density $N_e$ and the laser
amplitude $a_0$ change simultaneously, so that $S = N_e/a_0
N_c=const$, the laser-plasma dynamics remains similar. In particular,
this basic ultra relativistic similarity means that for different
interactions with the same $S=const$, the plasma electrons move
along similar trajectories while their momenta $\bf p$ scale as

\begin{equation} \label{SScalingG}
 {\bf p} \propto a_0.
\end{equation}

The $S-$similarity corresponds to a multiplicative transformation
group of the Vlasov-Maxwell equations, which appears in the
ultra relativistic regime. The similarity is valid for arbitrary
values of $S$. Physically the $S$-parameter separates relativistically
overdense plasmas ($S\gg1$) from underdense ones ($S\ll1$).

To apply the key result (\ref{SScalingG}) of the similarity theory
to the plasma surface motion, we rewrite (\ref{SScalingG}) for the
electron momentum components that are perpendicular $\textbf{p}_n$
and tangential $\textbf{p}_{\tau}$ to the plasma surface:

\begin{equation} \label{SScalingNT}
 {\bf p}_n\propto a_0,~~~~~~~~{\bf p}_{\tau}\propto a_0.
\end{equation}

\noindent
This result is significant. It shows that when we increase
the dimensionless vector potential $a_0$ of the incident wave while
keeping the plasma overdense, so that $S=const$, both $\textbf{p}_n$ and
$\textbf{p}_{\tau}$ grow as $a_0$. In other words the velocities of
the skin layer electrons

\begin{equation} \label{ElectronVelocity}
  v=c\sqrt{\frac{\textbf{p}_n^2+\textbf{p}_{\tau}^2}
  {m_e^2c^2+\textbf{p}_n^2+\textbf{p}_{\tau}^2}}=c(1-O(a_0^{-2}))
\end{equation}

\noindent
are about the speed of light almost at all times. Yet the
relativistic $\gamma$-factor of the plasma surface $\gamma_s(t')$
and its velocity $\beta_s(t')$ behave in a quite different way. To
realize this key fact let us consider the electrons at the very
boundary of the plasma. The scalings (\ref{SScalingNT}) state that
the momenta of these electrons can be represented as

\begin{eqnarray} \label{BoundaryMomenta}
  \textbf{p}_n(t')&=&a_0\textbf{P}_n(S,\omega_0 t') \\
  \textbf{p}_{\tau}(t')&=&a_0\textbf{P}_{\tau}(S,\omega_0 t'), \nonumber
\end{eqnarray}

\noindent
where $\textbf{P}_n$ and $\textbf{P}_{\tau}$ are universal
functions, which depend on the pulse shape and the $S$-parameter
rather than on $a_0$ or $N_e$ separately. Consequently, for
$\beta_s(t')$ and $\gamma_s(t')$ one obtains

\begin{eqnarray} \label{BoundaryDynamics}
  \beta_s(t')&=&\frac{p_n(t')}{\sqrt{m_e^2c^2+\textbf{p}_n^2(t')+\textbf{p}_{\tau}^2(t')}} \nonumber\\
             &=&\frac{P_n(t')}{\sqrt{\textbf{P}_n^2(t')+\textbf{P}_{\tau}^2(t')
  }} - O(a_0^{-2}),\\
  \gamma_s(t')&=&\frac{1}{\sqrt{1-\beta_s^2(t')}}
  =\sqrt{1+\frac{\textbf{P}_n^2(t')}{\textbf{P}_{\tau}^2(t')}}+O(a_0^{-2}).
\end{eqnarray}

One sees from (\ref{ElectronVelocity}) that when $a_0$ gets large,
the relativistic $\gamma$-factor of the electrons becomes large
too and their velocities approach the velocity of light. However,
the dynamics of the plasma boundary is significantly
different. For large $a_0$s the plasma boundary motion does not
enter the ultra relativistic regime and its relativistic
$\gamma$-factor $\gamma_s(t')$ is generally of the order of 
unity. Yet there is one exception: if at the moment $t'_g$ it
happens that $\textbf{P}_{\tau}(S,t'_g)=0$, i.e.

\begin{equation} \label{HarmonicsGeneration}
  \textbf{p}_{\tau}(S,t'_g)=0,
\end{equation}

\noindent we have

\begin{equation} \label{GammaJumps}
  \gamma_s=\frac{1}{\sqrt{1-\beta_s^2}}=\sqrt{\frac{\textbf{p}_n^2+m_e^2c^2}{m_e^2c^2}}\propto a_0.     
\end{equation}

\noindent
So the relativistic $\gamma$-factor of the boundary jumps to
$\gamma_s(t'_g)\propto a_0$ and the duration of the relativistic
$\gamma$-factor spike can be estimated as

\begin{equation} \label{Deltat}
 \Delta t'\propto 1/(a_0\omega_0).
\end{equation}

\noindent
For the velocity of the plasma boundary one finds analogously that
it smoothly approaches the velocity of light as
$\beta_s(t'_g)=(1-O(a^{-2}))$. Fig.~\ref{VelSurf} represents
schematically this behavior.

As we will see later, the $\gamma_s$ spikes cause the generation of
high harmonics in the form of ultra short laser pulses.


\section{Boundary Condition: Energy Conservation and the Apparent Reflection Point}
\label{BoundCond}


In this Section we introduce the boundary condition describing the
laser-overdense plasma interaction appealing to physical arguments,
just as it was previously done in \cite{Gordienko2004,Plaja1998}.
Mathematically rigorous analysis of the boundary condition is given
in \cite{TPBDiplomaThesis}. However, for the purposes of the present
work it is sufficient to treat this problem on a more intuitive
basis \cite{Gordienko2004}.

One might expect that the "oscillating mirror" model could describe
the laser-plasma interaction in our problem. Therefore we want to
explain in detail why it is not the case and then present the
derivation of the correct boundary condition.

The "oscillating mirror" model implies that the tangential components
of the vector potential are zero at the mirror surface. As a
consequence, if the ideal mirror moves with $\gamma\gg1$ towards
the laser pulse with the electric field $E_l$ and the duration
$\tau_0$, the reflected electric field will be
$E_{\mbox{\small refl}}\propto\gamma^2E_l$ and the pulse duration
will be $\tau_{\mbox{\small refl}}\propto\tau_0/\gamma^2$. The
energy of the reflected pulse would be then $\gamma^2$ times higher
than that of the incident one. However, since the plasma surface is
driven by the same laser pulse, this scaling is energetically
prohibited and consequently the plasma cannot be described as an
"ideal mirror".

To derive the correct boundary condition, let us consider the
tangential vector potential components of a laser pulse normally
incident onto a overdense plasma slab. These components satisfy
the equation

\begin{equation} \label{Maxwell}
 \frac{1}{c^2}\frac{\partial^2{\bf A}_{\tau}(t,x)}{\partial t^2}- \frac{\partial^2{\bf
  A}_{\tau}(t,x)}{\partial x^2}=\frac{4\pi}{c}{\bf j}(t,x),
\end{equation}

\noindent
where ${\bf A}_{\tau}(t,x=-\infty)=0$ and ${\bf j}$ is the
tangential plasma current density. Eq. (\ref{Maxwell}) yields

\begin{equation} \label{Green}
 {\bf A}_{\tau}(t,x)={2\pi}\int\limits_{-\infty}^{+\infty}{\bf J}\left(t,x,t',x'\right)\,dt'dx'. 
\end{equation}

\noindent
Here ${\bf J}\left(t,x,t',x'\right)={\bf
j}\left(t',x'\right) \left(\Theta_--\Theta_+\right)$, where we have
defined $\Theta_-=\Theta\left(t-t'-\left|x-x'\right|/c\right)$ and
$\Theta_+=\Theta\left(t-t'+\left(x-x'\right)/c\right)$, using the
Heaviside step-function $\Theta(t)$. Due to this choice of ${\bf J}$
the vector potential ${\bf A}_{\tau}(t,x)$ satisfies both
Eq. (\ref{Maxwell}) and the boundary condition at $x=-\infty$ since
${\bf J}(t,x=-\infty,t',x')=0$. The tangential electric field is
${\bf E}_{\tau} = -(1/c)\partial_t {\bf A}_{\tau}(t,x)$. If we
denote the position of the electron fluid surface by $X(t)$ we have

\begin{eqnarray} \label{Integrated}
 {\bf E}_{\tau}(t,X(t))=\frac{2\pi}{c}\sum\limits_{\alpha=-1}^{\alpha=+1}
  \alpha\int\limits_{0}^{-\infty}{\bf
  j}(t+\alpha\xi/c,X(t)+\xi)\,d\xi 
\end{eqnarray}

\noindent where $\xi = x' - X(t)$.

Now one has to estimate the parameters characterizing the skin
layer, i.e. the characteristic time $\tau_s$ of skin layer evolution
(in the co-moving reference frame) and the skin layer thickness
$\delta$.  Since the plasma is driven by the light pressure, one
expects that $\tau_s\propto 1/\omega_0$. The estimation of $\delta$
is more subtle.  From the ultra relativistic similarity theory
follows that $\delta\propto (c/\omega_0) S^{\Delta}$, where $S\gg 1$
for strongly overdense plasmas and $\Delta$ is an exponent that has
to be found analytically. In this work we do not discuss the exact
value of $\Delta$, but notice that this quantity does not depend on
neither $S$, nor $a_0$.  On the other hand, the denser the plasma,
the less the value of $\delta$.  This condition demands that
$\Delta<0$ and we get $c/\omega_0\gg \delta$ for $S\gg 1$.

If the characteristic time $\tau_s$ of the skin layer evolution is
long ($c\tau_s\gg \delta$), then we can use the Taylor expansion
${\bf j}\left(t\pm\xi/{c}, x'=X(t)+\xi\right)\approx {\bf
j}(t,x')\pm\epsilon$, where $\epsilon = (\xi/c)\partial_{t}{\bf
j}(t,x')$, and substitute this expression into (\ref{Integrated}).
The zero order terms cancel each other and we get
${\bf E}_{\tau}(t,X(t))\propto J_p (\delta/c\tau) \ll E_l$,
where $J_p\propto cE_l$ is the maximum plasma surface current. Thus,
as long as the skin-layer is thin and the plasma surface current is
limited, we can use the Leontovich boundary condition \cite{Landau8}

\begin{equation} \label{boundaryCondition}
 {\bf e}_n\times{\bf E}(t,X(t))=0. 
\end{equation}

\noindent
This condition has a straightforward relation to energy conservation.
Indeed, if we consider the Poynting vector

\begin{equation} \label{Pointing}
  \textbf{S}=\frac{c}{4\pi}\textbf{E}\times\textbf{B}.
\end{equation}

\noindent
we notice that the boundary condition (\ref{boundaryCondition})
represents balance between the incoming and reflected electromagnetic
energy flux at the boundary $X(t)$.

The boundary condition (\ref{boundaryCondition}) allows for another
interpretation. An external observer sees that the electromagnetic
radiation gets reflected at the point $x_\text{\tiny ARP}(t)$, where
the normal component of the Poynting vector
${\bf S}_n = c{\bf E}_\tau\times{\bf B}_\tau/4\pi= 0$, implied by
${\bf E}_\tau(x_\text{\tiny ARP})=0$. We call the point
$x_\text{\tiny ARP}(t)$ {\it the apparent reflection point} (ARP).

The actual location of the ARP can be
easily found from the electromagnetic field distribution in front
of the plasma surface. The incident laser field in vacuum runs in
the negative $x-$direction, ${\bf E}^i(x,t)={\bf E}^i(x+ct)$, while
the reflected field is translated backwards: ${\bf E}^r(x,t)={\bf E}^r(x-ct)$.
The tangential components of these fields interfere destructively
at the ARP position $x_\text{\tiny ARP}(t)$, so that the implicit
equation for the apparent reflection point $x_\text{\tiny ARP}(t)$
is:

\begin{equation} \label{ARP}
 {\bf E}_{\tau}^i(x_\text{\tiny ARP}+ct)+{\bf E}_{\tau}^r(x_\text{\tiny ARP}-ct) =0. 
\end{equation}

\noindent
We want to emphasize that Eq.~(\ref{ARP}) contains the electromagnetic
fields in vacuum. That is why the reflection point $x_\text{\tiny ARP}$
is {\it apparent}. The real interaction within the plasma skin layer
can be very complex. Yet, the external observer, who has information
about the radiation in vacuum only, sees that ${\bf E_\tau} = 0$ at
$x_\text{\tiny ARP}$. The ARP is located within the skin layer at the
electron fluid surface, which is much shorter than the laser wavelength
for overdense plasmas, for which the similarity parameter is $S\gg1$.
In this sense, the ARP is attached to the oscillating plasma surface.


\section{High Harmonic Universal Spectrum}
\label{SectionUniversalSpectrum}


According to Eq. (\ref{boundaryCondition}), the electric field of
the reflected wave at the plasma surface is

\begin{equation} \label{reflection}
 {\bf E}_r(t',X(t'))=-{\bf E}_i(t',X(t')),
\end{equation}

\noindent
where ${\bf E}_i(t', X(t'))=-(1/c)\partial_{t'}{\bf A}_i(t',X(t'))$
is the incident laser field, $t'$ is the reflection time. The
one-dimensional wave equation translates signals in vacuum without
change. Thus the reflected wave field at the observer position
$x$ and time $t$ is ${\bf E}_r(t,x)=-{\bf E}_i(t',X(t'))$. Setting
$x=0$ at the observer position we find that the Fourier spectrum of
the electric field ${\bf E}_r(t,x=0)$ is

\begin{eqnarray} \label{Spectrum}
 {\bf E}_r(\Omega)=\frac{m_ec\omega}{e\sqrt{2\pi}}
  \int\limits_{-\infty}^{+\infty} \text{Re}[i{\bf
  a}\left((ct'+X(t')\right)/c\tau_0)\times \nonumber\\
  \times \exp(-i\omega_0 t'- i\omega_0
  X(t')/c)]\exp(-i\Omega t)\,dt,
\end{eqnarray}

\noindent where

\begin{equation} \label{retardation}
 t'-X(t')/c=t
\end{equation}

\noindent
is the retardation relation \cite{Lichters1996}.

The fine structure of the spectrum of ${\bf E}_r(t)$ depends on the
particular surface motion $X(t)$, which is defined by the complex
laser- plasma interaction at the plasma surface.  Previous
theoretical works on high order harmonic generation from plasma surfaces
\cite{Lichters1996,Gibbon1996,vonDerLindeRzazewski1996,Plaja1998}
tried to approximate the function $X(t)$ in order to evaluate the
harmonic spectrum. For the first time analytical description of the
high harmonic intensity spectrum and the concept of universality
were presented in \cite{Gordienko2004}. This work proclaims the idea
that the most important features of the high harmonic spectrum do
not depend on the detailed structure of $X(t)$. However, the article
\cite{Gordienko2004} treats the universal spectrum without any
relation to similarity theory for ultra relativistic plasmas and, as
a result, relies on the saddle-point method without proper analysis
whether or not the areas contributing to the spectrum overlap
\cite{Gordienko2004,SaddlePoint}. In our analysis we overcome these
shortcomings of \cite{Gordienko2004}.

To find the spectrum, we notice that the investigation of ${\bf
E}_r(\Omega)$ (\ref{Spectrum}) is equivalent to the investigation of
the function

\begin{equation} \label{SpectrumModel}
 f(n)=f_{+}(n)+f_{-}(n),
\end{equation}

\noindent where

\begin{equation} \label{Example}
  f_{\pm}=\pm\int\limits_{-\infty}^{+\infty}\textbf{g}(\tau'+x(\tau'))
  \exp(\pm i(\tau'+x(\tau'))-in\tau)\,d\tau. 
\end{equation}

\noindent
Here $\tau=\omega_0 t$, $\tau'= \omega_0 t'$,
$n=\Omega/\omega_0$, $x(\tau')=(\omega_0/c)X(t')$ and $\textbf{g}$
is a slowly varying function $\left(|dg(\tau')/d\tau'|\ll1\right)$,
which is trivially related to $\textbf{a}$ as

\begin{equation} \label{RelationGandA}
  \textbf{g}(\tau'+x(\tau'))=\frac{-im_ec}{2e\sqrt{2\pi}}
  \textbf{a}((ct'+X(t'))/c\tau_0).
\end{equation}

\noindent
Making use of Eq. (\ref{retardation}) we re-write Eq. (\ref{Example}) as

\begin{eqnarray} \label{Example1}
  f_{\pm}=\pm\int\limits_{-\infty}^{+\infty}g(\tau'+x(\tau'))
  \exp(i\tau'(-n\pm1)+  \nonumber\\
  +~ix(\tau')(n\pm1))\left(1-x'\left(\tau'\right)\right)\,d\tau'.
\end{eqnarray}

\begin{figure}
  \centerline{\includegraphics[width=7cm,clip]{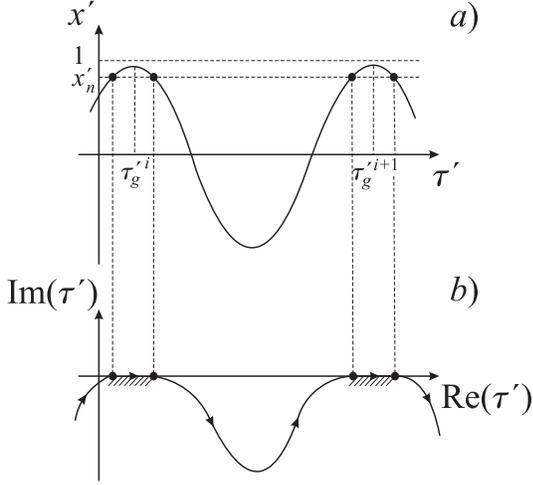}}
 \caption{Surface dynamics and path integration in
  (\ref{Example1}). a) Velocity $x'(\tau')$ of the plasma
  surface; $x'_n = (n-1)/(n+1)$ are the saddle points corresponding to
  $d\Theta/d\tau'=0$. b) The integration path can be shifted below
  the real axis everywhere except in the neighborhoods of $\tau'_g$
  (dashed regions).} 
  \label{Velocity}
\end{figure}

\noindent
We wish to examine the integral (\ref{Example1}) for very large $n$. 
For this purpose, we notice that the derivative of the phase

\begin{equation} \label{Phase}
  \Theta(\tau')=\tau'(-n\pm1)+x(\tau')(n\pm1)
\end{equation}

\noindent
is negative everywhere except in the vicinity of
$\tau'_g=\omega_0 t'_g$ for which $x'(\tau'_g)\approx 1$ (see
Fig.~\ref{Velocity}a).

The physical meaning of $\tau'_g$ and the behavior of $x(\tau')$ in the
vicinity of these times is explained by
Eq. (\ref{HarmonicsGeneration}). Since the time derivative of
$\Theta(\tau')$ is negative for all $\tau$s that are not too close
to one of the $\tau'_g$, we can shift the path over which we
integrate to the lower half of the complex plane everywhere except
in the neighborhoods of $\tau'_g$ (see Fig.~\ref{Velocity}b).  The
contributions of the parts remote from the real axis are
exponentially small. We can shift the path to the complex plane till
the derivative equals zero or we find a singularity of the phase
$\Theta$.

To calculate the contributions of $\tau'_g$s neighborhoods we can
expand $x'(\tau')$ near each of its maxima at $\tau'_g$. Since every
smooth function resembles a parabola near its extrema, the expansion
of $x'(\tau')$ is a quadratic function of $(\tau'-\tau'_g)$. Simple
integration leads to the following expression for $x(\tau')$

\begin{equation} \label{Expansion}
  x(\tau')=x(\tau'_g)+v_0(\tau'_g)(\tau'-\tau'_g)-
  \frac{\alpha(\tau'_g)}{3}(\tau'-\tau'_g)^3. 
\end{equation}

\noindent
The Taylor expansion given by Eq. (\ref{Expansion}) has three
important properties related to its dependence on $S$ and $a_0$:
1) for $S=const$ and $a_0\to+\infty$ one finds that
   $v_0\to c$;
2) for $a_0\to+\infty$, $\alpha$ depends only
   on the parameter $S$;
3) the expansion (\ref{Expansion}) is
   a good approximation for $\left|\tau'-\tau'_g\right|\ll
   (2\pi/\omega_0)f_1(S)$, where the function $f_1$ does not
  depend on $a_0$.
These three properties are mathematical statements of the
physical picture described in Section \ref{PhysPict} combined with
the similarity theory developed in Section \ref{KinDescr}. In other
words, the properties of the expansion (\ref{Expansion}) just
mentioned are direct consequences of the physical picture presented
in Fig.~\ref{VelSurf}.

Substitution of Eq. (\ref{Expansion}) into $f_{\pm}(n)$ yields

\begin{equation} \label{Contributions}
  f_{\pm}(n)=\sum_{\tau'_g}f_{\pm}(\tau'_g,n),
\end{equation}

\noindent where the sum is over all times $\tau'_g$,

\begin{eqnarray}
 \label{spectrumDetails1}
    f_{+}(\tau'_g,n)&=&g\left(\tau'_g+x(\tau'_g)\right)\exp(i\Theta_{+}(\tau'_g,n))\times \nonumber\\
    & &\times F(\tau'_g,n)\\
 \label{spectrumDetails2}
    f_{-}(\tau'_g,n)&=&-g\left(\tau'_g+x(\tau'_g)\right)\exp(i\Theta_{-}(\tau'_g,n))\times \nonumber\\
    & &\times F(\tau'_g,-n)\\ \label{Airy}
    F(\tau'_g,n)&=&\frac{4\sqrt{\pi}}{{\alpha(\tau'_g)}^{\frac{1}{3}}n^{\frac{4}{3}}}
       Ai\left(\frac{2}{n_{cr}(\tau'_g)}\frac{n-n_{cr}(\tau'_g)}{(\alpha(\tau'_g)n)^{1/3}}\right)\\
  \label{spectrumDetails3}
    \Theta_{\pm}&=&\pm(\tau'_g+x(\tau'_g))+n(x(\tau'_g)-\tau'_g),
\end{eqnarray}

\noindent
and $n_{cr}=2/(1-v_0)$. In (\ref{Airy}) $Ai$ is the well-known
Airy-function, defined as

\begin{equation} \label{AiryDefinition}
 Ai(x)=\frac{1}{\sqrt{\pi}}\int\limits_{0}^{+\infty}\cos\left(ux+\frac{1}{3}u^3\right)du.
\end{equation}

\noindent
Note that if $x(\tau'+\pi)=x(\tau')$ and $g(\tau')=g(\tau'+\pi)$,
then $f_{\pm}(2n)=0$.

Using equations (\ref{Contributions})-(\ref{spectrumDetails3}) we
can show analytically that the spectrum of radiation generated by the
plasma is described by a universal formula.

For the intensity of the $n$th harmonic we obtain

\begin{equation} \label{Intensity}
  I_n\propto\left|\exp\left(i\Theta_{+}(n)\right)F\left(n\right)
  -\exp\left(i\Theta_{-}(n)\right)F\left(-n\right)\right|^2,
\end{equation}

\noindent where

\begin{eqnarray}
  F(n)&=&\frac{4\sqrt{\pi}}{(\sqrt[4]{\alpha}n)^{4/3}}Ai\left(\frac{2}{n_{cr}}
  \frac{n-n_{cr}}{(\alpha n)^{1/3}}\right),\\
  \Theta_{\pm}(n)&=&\pm\Theta_0-n\Theta_1,
\end{eqnarray}

\noindent
with the Airy function $Ai(x)$ defined in (\ref{AiryDefinition})
and the critical harmonic number $n_{cr}$ satisfying $n_{cr}=4\gamma_{\max}^2$,
where $\gamma_{\max}$ is the largest relativistic factor of the
plasma boundary.

Eq. (\ref{Intensity}) gives an exact formula for the high harmonic
spectrum, which includes both power law and exponential decay parts.
Now we want to use different asymptotic representations of the Airy
function in order to demonstrate explicitly these two quite different
laws of high harmonic intensity decay.

For $n<\sqrt{\alpha/8}n_{cr}^{3/2}$ ($2\left|1-n/n_{cr}\right|\ll(\alpha
n)^{1/3}$) we can substitute the value of the Airy function at $x=0$
($Ai(0)=\sqrt{\pi}/(3^{2/3}\Gamma(2/3))=0.629$,
$Ai'(0)=-3^{1/6}\Gamma(2/3)/(2\sqrt{\pi})=-0.459$) in
Eq. (\ref{Intensity}), and obtain

\begin{equation} \label{8/3}
  I_n\propto\frac{1}{n^{8/3}}\left|\sin\Theta_0
  +\frac{Ai'(0)}{Ai(0)}B(n,\Theta_0)\right|^2,
\end{equation}

\noindent where

\begin{equation} \label{Correction}
  B(n,\Theta_0)=\frac{2\sin\Theta_0}{(\alpha
  n_{cr})^{1/3}}\left(\frac{n}{n_{cr}}\right)^{2/3}+
  \frac{2i\cos\Theta_0}{(\alpha n)^{1/3}}. 
\end{equation}

\noindent
This means that the universal spectrum

\begin{equation} \label{Universal}
 I_n\propto 1/n^{8/3}
\end{equation}

\noindent
is observed everywhere except for $\sin\Theta_0\approx 0$, when
the dominant term in the expansion is zero. For this particular
case, a higher order correction is important for

\begin{equation} \label{universalCondition}
  (\alpha n)^{1/3}\tan\Theta_0<2\left|\frac{Ai'(0)}{Ai(0)}\right|
\end{equation}

\noindent
and in this restricted frequency range the spectrum

\begin{equation} \label{Spectrum10/3}
 I_n\propto 1/n^{10/3}
\end{equation}

\noindent should be used.

At this point we want to explain the meaning of $8/3$-spectrum
universality. Notice that since Eq. (\ref{8/3}) depends on the
phase $\Theta$, for moderate values of $n_{cr}$ the best power law
fit of the high harmonic spectrum can be delivered by

\begin{equation} \label{bestFit}
  I_n\propto n^{-p},
\end{equation}

\noindent
where $8/3\le p\le10/3$. When $n_{cr}$ becomes really large, the
majority of the harmonics does not satisfy the inequality (\ref{universalCondition})
and the spectrum inevitably becomes $I_n\propto n^{-8/3}$. In
other words, one should think of the $8/3$-spectrum as the high
intensity limit of the high harmonic spectrum. To find an
analytical criterion for the $8/3$-spectrum generation one can
state that the condition (\ref{universalCondition}) has to be
violated for the harmonics with $n\propto\sqrt{\alpha/8}n_{cr}^{3/2}$.
This means that

\begin{equation} \label{gamma8/3}
  \gamma_{max}>\gamma_{8/3}=\frac{1}{\sqrt{2}\tan{\Theta_0}}\left|\frac{Ai'(0)}{Ai(0)}\right|.
\end{equation}

The formal expansion of the Airy function for $n\ll n_{cr}$,
$(\alpha n)^{1/3}<1$ leads to the spectrum $I_n\propto1/n^{5/2}$
discussed in \cite{Gordienko2004}. Meticulous analysis
demonstrates that this case is irrelevant. Yet the version of
the saddle point method used in \cite{Gordienko2004} ascribes the
spectrum $1/n^{5/2}$ to the whole area below the cutoff. This is
because the areas around the saddle points contributing to the
spectrum overlap significantly and the approximation used in
\cite{Gordienko2004} is not accurate enough in this part of the
spectrum. However, the difference between the powers $8/3$ and
$5/2$ is only $1/6$ and it is hardly distinguishable numerically
and experimentally.

For $n>\sqrt{\alpha/8}n_{cr}^{3/2}$ (2$\left|1-n/n_{cr}\right|\gg(\alpha
n)^{1/3}$) Eq. (\ref{Airy}) can be rewritten as

\begin{equation} \label{TheTail}
 I_n\propto\frac{n_{cr}^{1/2}}{n^3}\exp\left(-\frac{16\sqrt{2}}{3\alpha^{1/2}}
  \frac{n-n_{cr}}{n_{cr}^{3/2}}\right). 
\end{equation}

\noindent
It is interesting to notice that the approximation used in
\cite{Gordienko2004} also gives Eq. (\ref{TheTail}) for this area,
i.e. here the overlapping of the contributing areas is really
negligible.


\section{Ultra short pulse duration}

For ultra short pulse generation not only the amplitudes but also
the harmonic phases are of importance. The calculations presented
above show that all harmonics are phase locked. This means that
after proper filtering they can produce a pulse of duration $T$,
such that

\begin{equation} \label{Duration}
  T\propto\frac{\pi}{\omega_0}\frac{1}{\sqrt{\alpha}n_{cr}^{3/2}}
  \propto\frac{1}{\sqrt{\alpha}\gamma_{\max}^3},
\end{equation}

\noindent
where $\omega$ is the frequency of the fundamental wave.

Eq.~(\ref{Duration}) presents a new result. Notice that the plasma
boundary never moves with a relativistic $\gamma$-factor larger than
$\gamma_{\max}$. Consequently, the frequency of a photon reflected
from this boundary due to the relativistic Doppler effect does not
exceed $4\gamma_{\max}^2\omega_0=n_{cr}\omega_0$. How can a pulse with
duration $T$ given by Eq. (\ref{Duration}) be produced then?

Mathematically this can be understood looking at the properties of
the Airy function. $Ai(x)$ changes its behavior from oscillatory for
$x<0$ with $|x|\gg 1$

\begin{equation} \label{Oscillatory}
  Ai(x)\approx\frac{1}{|x|^{1/4}}\sin\left(\frac{2}{3}|x|^{3/2}+\frac{\pi}{4}\right)
\end{equation}

\noindent to exponentially decaying for $x\gg1$

\begin{equation} \label{Exponential}
 Ai(x)\approx\frac{1}{2x^{1/4}}\exp\left(-\frac{2}{3}x^{3/2}\right).
\end{equation}

\noindent
The point $x=0$ corresponds to $n=n_{cr}$. However, the exponential is
so small for $n<\sqrt{\alpha/8}n_{cr}^{3/2}$ that the power law decay
dominates over the exponent. As a result, the power law spectrum
goes well beyond the threshold $\omega_0n_{cr}$ predicted by
oversimplified models considering reflection from a moving
surface.

The reasoning just presented explains the mathematical origin of the
$\sqrt{\alpha/8}n_{cr}^{3/2}$-cutoff. In the next section we give a
simple physical interpretation of this result and reveal its relation
to the relativistic $\gamma$-factor spikes. We also take a closer
look at the mathematical nature of the cutoff and its relation to
the failure of the standard saddle point approach due to
saddle point overlapping.


\section{Relativistic Spikes and Cut-Off of the High Harmonic Spectrum}
\label{SaddlePoints}


The $\gamma^3-$scaling of the spectrum cutoff (\ref{ap:cutoff}) is
readily understood using the  relativistic $\gamma$-factor spikes of
the plasma surface motion. Indeed, the plasma surface velocity in the
vicinity of a maximum can be approximated as

\begin{equation} \label{velocityExpansion}
  v(t)=v_0(t_g)-\alpha\omega_0^2(t-t_g)^2
\end{equation}

\noindent
(see Section \ref{SectionUniversalSpectrum}). Consequently, for the surface
$\gamma$-factor during a relativistic spike we find

\begin{equation} \label{gammaFactorExpansion}
  \gamma(t)\approx\frac{\gamma_{\max}}{\sqrt{1+\gamma^2_{\max}\alpha\omega_0^2(t-t_g)^2}},
\end{equation}

\noindent
where $\gamma_{\max}=1/\sqrt{1-v_0^2(t_g)/c^2}$. Eq. (\ref{gammaFactorExpansion})
shows that the highest harmonics are generated over the time interval

\begin{equation} \label{timeInterval}
 \Delta t\propto\frac{1}{\omega_0}\frac{1}{\sqrt{\alpha}\gamma_{\max}}.
\end{equation}

\noindent
For the whole time interval $\Delta t$ the relativistic spike moves
with ultra relativistic velocity in the direction of the emitted radiation. 
For this reason, the spatial length $L$ of the high harmonic pulse produced is

\begin{equation} \label{pulseLength}
 L\propto (c-v_0(t_g))\Delta t\propto \frac{c}{\omega_0}\frac{1}{\sqrt{\alpha}\gamma_{\max}}.
\end{equation}

\noindent
A pulse of such duration contains frequencies

\begin{equation} \label{frequences}
 \Omega\propto \frac{c}{L}\propto\omega_0\sqrt{\alpha}\gamma_{\max}^3,
\end{equation}

\noindent
what physically explains the origin of the high frequency
cutoff (\ref{ap:cutoff}). This cutoff should be compared
with the one predicted by the "oscillating mirror" model: 
$4\gamma_{\max}^2$. It differs parametrically from the
correct cutoff $\propto\sqrt{8\alpha}\gamma_{\max}^3$, which is due to
the relativistic $\gamma$-factor 
spikes. 

As it has been shown, the "oscillating mirror" model gives the
incorrect formula for the spectrum cutoff because it does not include
the relativistic $\gamma$-factor spikes. 
Mathematically, this failure "oscillating mirror" model is related to
the saddle point overlapping. Eq. (\ref{Phase}) defines the saddle 
points

\begin{equation} \label{saddlePoints}
  \frac{dx(\tau')}{d\tau'}=1\pm\frac{1}{n},
\end{equation}

\noindent
the vicinity of which determines the value of the integral
describing the amplitude of the $n$-th harmonic, $n\gg1 $, see
Fig.~\ref{Velocity}. Eq.~(\ref{Expansion}) yields that the set of
Eqs.~(\ref{saddlePoints}) 
has real (not imaginary) solutions only for $n<n_{cr}=4\gamma_{\max}^2$.
For larger $n$ all of the saddle points of (\ref{saddlePoints}) have
an imaginary part.

Let us now apply the standard saddle point method to the problem
without investigating whether the conditions for its applicability
are met (it is clear that these conditions are violated at least for
$n\approx n_{cr}$, for which two saddle points coincide). This
approximation predicts that for $1\ll n \ll n_{cr}$ the spectrum
decays as $1/n^{5/2}$.  For $n>n_{cr}$ this approach restores
Eq. (\ref{Exponential}).

As it was mentioned above the spectrum $1/n^{5/2}$ occurs formally
in the limit $(\alpha n)^{1/3}<1$. In other words, if $\alpha$ were
small, this spectrum would be observed. However, since $\alpha$
depends on $S$ rather than $a_0$, this spectrum corresponds to
harmonics with small numbers only and is hardly of any practical
interest. On the other hand, one can notice that $\alpha$ describes
the plasma surface acceleration at the maximum of the velocity. This
means that the limit of small $\alpha$ (and the spectrum $1/n^{5/2}$)
describes the limit of the surface moving without acceleration.
However, this limit is of little interest for large $a_0$.


\section{Cutoff and the structure of filtered pulses}
\label{PulseStructure}


As we have seen, the relativistic plasma harmonics are phase locked
and can be used to generate ultra short pulses. However, to extract
these ultra short pulses, one has to remove the lower harmonics. 
The high energy cutoff (\ref{ap:cutoff}) of the power law spectrum 
defines the shortest pulse duration that can be achieved this way.

Let us apply a high-frequency filter that suppresses 
all harmonics with frequencies below $\Omega_f$ and study 
how the relative position of the $\Omega_f$ and the spectrum cutoff
affects the duration of the resulting (sub)attosecond pulses.

According to Eq. (\ref{Intensity}), the electric field of the pulse
after the filtration is

\begin{eqnarray} \label{Filtration}
 E\propto Re\int\limits_{\Omega_f/\omega_0}^{+\infty}(\exp\left(i\Theta_{+}(n)\right)F(n)
  -\exp\left(i\Theta_{-}(n)\right)\times& & \nonumber\\
  \times F(-n))\exp(int)\,dn& &.
\end{eqnarray}

The structure of the filtered pulses depends on where we set the
filter threshold $\Omega_f$. In the case
$1\ll\Omega_f/\omega_0\ll\sqrt{\alpha/8}n_{cr}^{3/2}$, we use
Eq. (\ref{Universal}) and rewrite
Eq. (\ref{Filtration}) as

\begin{eqnarray} \label{LowFrequency}
  E\propto
  Re\int\limits_{\Omega_f/\omega_0}^{+\infty}\frac{\exp(int)}{n^{4/3}}\,dn=~~~~~~~~~~~~~~~~~~~~~~~~&& \nonumber\\
  =\left(\frac{\omega_0}{\Omega_f}\right)^{1/3}Re\exp(i\Omega_ft-i\Theta_1)P(\Omega_ft),&&
\end{eqnarray}

\noindent where the function $P$

\begin{equation} \label{UniversalFunction}
  P(x)=\int\limits_1^{+\infty}\frac{\exp(iyx)}{y^{4/3}}\,dy
\end{equation}

\noindent
gives the slow envelope of the pulse. 

It follows from the expression (\ref{LowFrequency}) that
the electric field of the filtered pulse decreases very slowly with
the filter threshold 
as $\Omega_f^{-1/3}$. The pulse duration decreases as
$1/\Omega_f$. At the same time, the fundamental frequency of the pulse is 
$\Omega_f$. Therefore the pulse is hollow when $\Omega_f/\omega_0
\ll\sqrt{\alpha 8}\gamma_{\max}^{3}$, 
i.e. its envelope is not filled with electric field oscillations. One
possible application of these pulses 
is to study atom excitation by means of a single strong kick.

The pulses structure changes differ significantly when the filter
threshold is placed above the spectrum cutoff. For
$\Omega_f/\omega_0\gg \sqrt{\alpha 8}\gamma_{\max}^{3}$ we use
Eqs. (\ref{TheTail}) and (\ref{Filtration}) to obtain

\begin{eqnarray} \label{HighFrequency}
 E\propto\left(\frac{\omega_0}{\Omega_f}\right)^{3/2}
  \exp\left(-\frac{8\sqrt{2}\Omega_f}{3\sqrt{\alpha}\omega_0n_{cr}^{3/2}}\right)
  \times&&~~~~~~~~~~~~~~ \nonumber\\
  \times Re\frac{\exp(i\Omega_f t-i\Theta_1)}{8\sqrt{2}/\left(3\sqrt{\alpha}n_{cr}^{3/2}\right) +
  i\omega_0 t}.&& 
\end{eqnarray}

\noindent
Th amplitude of these pulses decreases fast when
$\Omega_f$ grows. However, the pulse duration
$\propto 1/\sqrt{\alpha}\gamma_{\max}^{3}$ does 
not depend on $\Omega_f$. Since the fundamental frequency of the pulse
grows as $\Omega_f$, the pulses obtained with an
above-cutoff filter are filled with electric field
oscillations. Therefore these pulses are suitable to study the resonance
excitation of ion and atom levels.

The minimal duration of the pulse obtained by cutting-off low order harmonics
is defined by the spectrum cutoff $\propto \sqrt{\alpha 8}\gamma_{\max}^{3}$.
Physically, this result is the consequence of the ultra relativistic
spikes in the plasma surface $\gamma-$factor.


\section{Spectrum Modulations}
\label{SpectrumModulations}


In this section we discuss the harmonic phases and show how the
interference of harmonics produced by different $\gamma-$spikes can
lead to spectrum modulations.

Eq. (\ref{Contributions}) obtained in Section
\ref{SectionUniversalSpectrum} allows for straightforward physical
interpretation. The sum $f_{-}(\tau'_g,n)+f_{+}(\tau'_g,n)$ gives
the contribution of the $\tau'_g$-spike of the surface relativistic
factor to the harmonic spectrum (see Fig.~\ref{Velocity}).
Therefore the phase of the $n$th harmonic $\phi_n$ due to the
$\tau'_g$s spike is given by

\begin{eqnarray} \label{Phase1}
  \tan(\phi_n(\tau'_g)&+&n\Theta_1(\tau'_g))
  \tan\left(\Theta_0\right)=\nonumber\\
  &=&\frac{1-F(\tau'_g,-n)/F(\tau'_g,n)}{1+F(\tau'_g,-n)/F(\tau'_g,n)},
\end{eqnarray}

\noindent
where $\Theta_0(\tau'_g)=\tau'_g+x(\tau'_g)$ is the phase of the
incident laser pulse at the time $\tau'_g$ and
$\Theta_1(\tau'_g)=\tau'_g-x(\tau'_g)=\tau(\tau'_g)$, the time of
the observation, i.e. the symbol $\Theta_1$ is redundant, yet we
keep it for the sake of notation coherency.

Since $F(\tau'_g,-n)\approx F(\tau'_g,n)$ for $n\gg 1$,
Eq. (\ref{Phase1}) gives rise to $\phi_n(\tau'_g)\approx
-n\Theta_1(\tau'_g)$. This means that each spike radiates phase
locked high harmonics that can be used for ultra short pulse
production. Another consequence of $F(\tau'_g,-n)\approx
F(\tau'_g,n)$ is that the amplitude of the harmonics is proportional
to $\sin\left(\Theta_0(\tau'_g)\right)$.

The mechanism presented in Fig.~\ref{VelSurf} has another very
interesting consequence. Each harmonic is generated due to several
spikes.  These spikes contribute to (\ref{Contributions}) with
different phase multipliers.  This phase difference leads to
modulations in the high harmonic spectrum.  As an example we
consider the interference between the harmonics produced by two
different spikes in detail. The phase shift between the
contributions from different spikes is
$\phi_n(\tau'_{g_1})-\phi_{n}(\tau'_{g_2})=
\Theta_0(\tau'_{g_1})-\Theta_0(\tau'_{g_2})
-n(\Theta_1(\tau'_{g_1})-\Theta_1(\tau'_{g_2}))$ and can be large
for large $n$. Since for
$n\ll\min\left(\sqrt{\alpha(\tau'_{g_1})/8}n_{cr}^{3/2}(\tau'_{g_1}),
\sqrt{\alpha(\tau'_{g_2})/8}n_{cr}^{3/2}(\tau'_{g_2})\right)$ only
$Ai(0)$ enters $f_{\pm}(\tau'_{g_{1,2}})$, the values of the
contributions from the $\tau'_{g_{1,2}}$ spikes for this harmonic
range do not depend on $n_{cr}(\tau'_{g_{1,2}})$. As a result the
modulations in this harmonic range depend only on the parameter
$S$.

To recapitulate, the non-trivial plasma motion producing more than one
$\gamma$-spike per oscillation period is the cause of the spectrum
modulation.


\section{Numerical results}
\label{NumRes}


In order to check our analytical results, we have performed a
number of 1d PIC simulations with the 1d particle-in-cell code
VLPL \cite{vlpl}. In all simulations a laser pulse with the
Gaussian envelope $a=a_0\exp\left(-t^2/\tau_L^2\right)\cos\left(\omega_0 t\right)$,
duration $\omega_0\tau_L=4\pi$ and dimensionless vector potential
$a_0=20$ was incident onto a plasma layer with a step density
profile.

\subsection{Apparent Reflection Point}

First, we study the oscillatory motion of the plasma and the dynamics
of the apparent reflection point defined by the boundary condition
(\ref{boundaryCondition}) \cite{Baeva2006}. The plasma slab
is initially positioned between $x_R=-1.5\lambda$ and $x_L=-3.9\lambda$,
where $\lambda = 2\pi/\omega_0$ is the laser wavelength. The laser pulse
has the amplitude $a_0=20$. The plasma density is $N_e/N_c=90$
($S=4.5$).

\begin{figure} [h]
  \centerline{\includegraphics[width=7.5cm,clip]{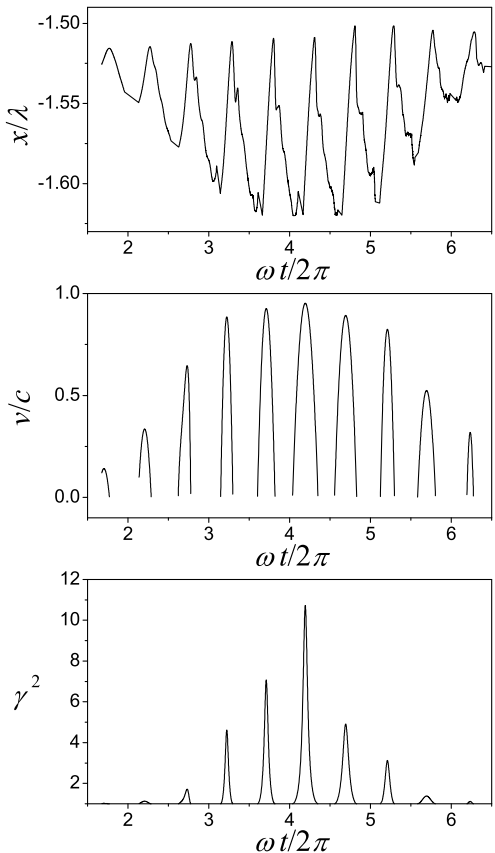}}
 \caption{1D PIC simulation results for the
           parameters $a_0=20$ and $N_e=90N_c$. a) Oscillatory motion of the
           point $x_\text{\tiny ARP}(t)$ where ${\bf E}_\tau(x(t))=0$.b)
           Velocity $v_\text{\tiny ARP}(t) = dx_\text{\tiny ARP}(t)/dt$; only
           the positive velocities are shown, since they correspond to motion
           towards the laser pulse in the geometry of this simulation.
           Notice that the ARP velocity is a smooth function. c) The
           corresponding $\gamma-$factor $\gamma_\text{\tiny ARP}(t) =
           1/\sqrt{1-v_\text{\tiny ARP}(t)^2/c^2}$ contains
           sharp spikes, which coincide with the velocity extrema.}
  \label{a20x_v_gamma}
\end{figure}

At every time step, the incident and the reflected fields are
recorded at $x=0$ (the position of the "external observer"). Being
solutions of the wave equation in vacuum, these fields can be easily
chased to arbitrary $x$ and $t$. To find the ARP position
$x_\text{\tiny ARP}$, we solve numerically equation (\ref{ARP}).
The trajectory of $x_\text{\tiny ARP}(t)$ obtained in this
simulation is presented in Fig.~\ref{a20x_v_gamma}a. One can
clearly see the oscillatory motion of the point $x_\text{\tiny
ARP}(t)$.  The equilibrium position is displaced from the initial
plasma boundary position $x_R$ due to the mean laser light pressure.

Since only the ARP motion towards the laser pulse is of importance
for the high harmonic generation, we cut out the negative ARP
velocities $v_\text{\tiny ARP}(t) = dx_\text{\tiny ARP}(t)/dt$ and
calculate only the positive ones (Fig.~\ref{a20x_v_gamma}b). The
corresponding $\gamma$-factor $\gamma_\text{\tiny
ARP}(t)=1/\sqrt{1-v_\text{\tiny ARP}(t)^2/c^2}$ is presented in
Fig.~\ref{a20x_v_gamma}c. Notice that the ARP velocity is a smooth
function. At the same time, the $\gamma$-factor $\gamma_\text{\tiny
ARP}(t)$ contains sharp spikes, which coincide with the velocity
extrema. These numerical results confirm the predictions of the
ultra relativistic similarity theory, which were presented in
Section \ref{KinDescr}.

\subsection{High harmonic spectrum}

\begin{figure} [h]
  \centerline{\includegraphics[width=8cm,clip]{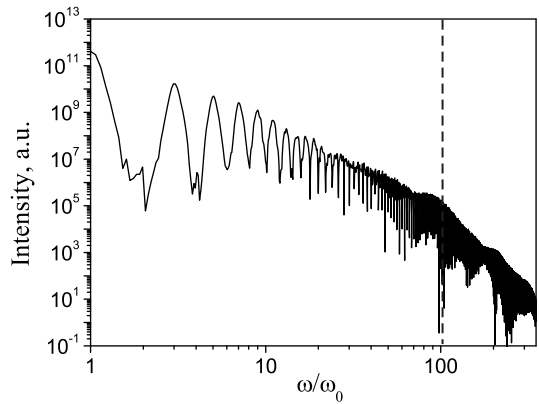}}
 \caption{Spectrum of high harmonics obtained numerically for the
           case of $a_0=20$ and $N_e=90N_c$, corresponding to
           $S=4.5$ and $\gamma_{max}\approx 3.3$. Assuming
  $\alpha\approx1$, the cutoff (1) is expected at $n\approx 100$. This
  analytically predicted cutoff is marked by the dashed line. }
 \label{spectrum2090}
\end{figure}

For the same laser-plasma parameters ($a_0=20$, $N_e=90N_c$) the
spectrum of high harmonic radiation is presented in Fig.~\ref{spectrum2090}.
The maximum $\gamma-$factor of the apparent reflection point in
this numerical simulation is $\gamma_{\max}\approx 3.3$ (compare
with Fig.~\ref{a20x_v_gamma}).
Consequently the maximal harmonic number predicted by the
"oscillating mirror" model lies at $4\gamma_{max}^2\approx 40$,
while the harmonic cutoff predicted by the relativistic spikes is
about 100. Fig.~\ref{spectrum2090} clearly demonstrates that there
is no change of the spectrum behavior at $4\gamma_{max}^2$, while 
steeper decay takes place above 100, as predicted by our theory.
Also, the spectral intensity modulations discussed in Section
\ref{SectionUniversalSpectrum} and \cite{Watts2002,Teubner2003}
are observed.

\begin{figure}
  \centerline{\includegraphics[width=8cm,clip]{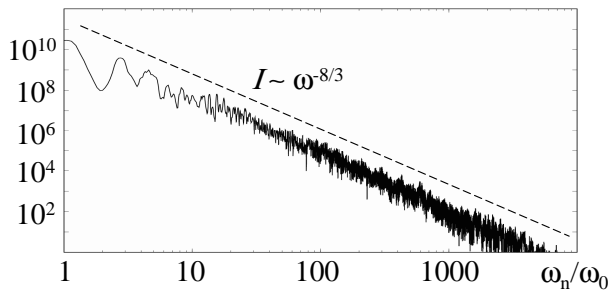}}
 \caption{Spectra of the reflected radiation for the
           laser amplitude $a_0=20$ and the plasma density $N_e=30
  N_{cr}$. The broken line marks the universal scaling $I\propto \omega^{-8/3}$.}  
 \label{figSpectra}
\end{figure}

To be able to make a real statement about the power in the power law
decay of the spectrum we need more harmonics in order to satisfy the
condition of universal $8/3$-spectrum formation (\ref{gamma8/3}). For
this reason we made the simulation with parameters $a_0=20$ and $N_e=30N_c$,
which roughly corresponds to solid hydrogen or liquid helium. The
reflected radiation spectrum obtained for these parameters is shown in 
Fig.~\ref{figSpectra} in log-log scale. 
The power law spectrum $I_n\propto 1/n^{8/3}$ is clearly seen here,
thus confirming the analytical results of Section
\ref{SectionUniversalSpectrum}. 

\subsection{Subattosecond pulses}

\begin{figure} [h]
  \centerline{\includegraphics[width=7cm,clip]{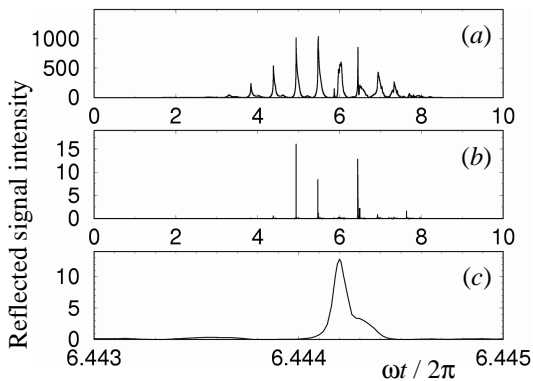}}
 \caption{Zeptosecond pulse train: a) temporal
           structure of the reflected radiation; b) zeptosecond pulse train
           seen after spectral filtering; c) one of the zeptosecond pulses
           zoomed, its FWHM duration is about 300~zs.} 
 \label{figZepto}
\end{figure}

Let us take a closer look at Fig.~\ref{figSpectra}. The power law
spectrum extends at least till the harmonic number 2000, and zeptosecond
($1\text{zs}=10^{-21}\text{s}$) pulses can be generated.  
The temporal profile of the reflected radiation
is shown in Fig.~\ref{figZepto}. When no spectral filter is applied,
Fig.~\ref{figZepto}a, a train of attosecond pulses is observed
\cite{Plaja1998}. 

\noindent
However, when we apply a spectral filter selecting
harmonics above $n=300$, a train of much shorter pulses is obtained,
Fig.~\ref{figZepto}b. Fig.~\ref{figZepto}c zooms in to one of these
pulses. Its full width at half maximum is about $300~$zs. At the same
time its intensity normalized to the laser frequency is huge
$(eE_{zs}/mc\omega)^2 \approx 14$. This corresponds to the
intensity $I_{zs}\approx 2\times10^{19}$~W/cm$^2$.

\subsection{Filter threshold and the attosecond pulse structure}

The dependence of the short pulses on the position of the filter also
can be studied numerically. We apply a filter with the filter function
$f(\omega)=1+\tanh((\omega-\Omega_f)/\Delta\omega)$. It passes through
frequencies above $\Omega_f$ and suppresses lower frequencies. We
choose the similation case of laser
vector potential $a_0=20$ and plasma density $N_e=90N_c$. The spectrum
of high harmonics is given in Fig.~\ref{spectrum2090}. 

\begin{figure} [h]
  \centerline{\includegraphics[width=9.5cm,clip]{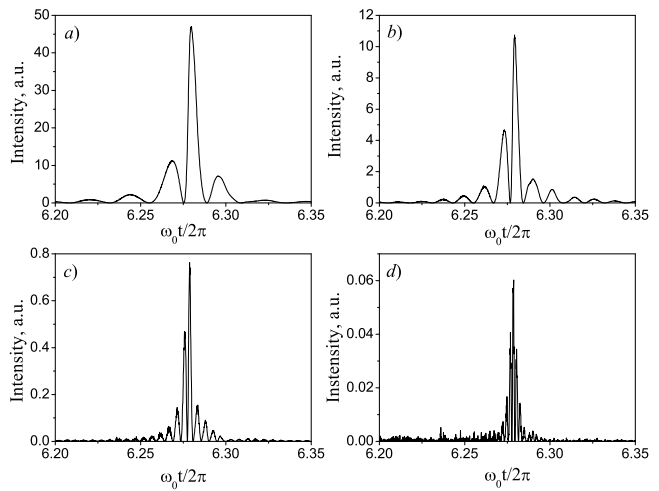}}
 \caption{Dependence of the pulses filling on the position of the
           sharp filter boundary for $a_0=20$ and $N_e=90N_c$ and
           filter positions: a) $\Omega_f=20\omega_0$, $\Delta\omega=2\omega_0$;
           b) $\Omega_f=40\omega_0$, $\Delta\omega=2\omega_0$;
          c) $\Omega_f=100\omega_0$, $\Delta\omega=2\omega_0$;
           d) $\Omega_f=200\omega_0$, $\Delta\omega=2\omega_0$}
 \label{shortPulses}
\end{figure}

\noindent
We zoom in to one of the pulses in the pulse train obtained and
study how the shape of this one pulse changes with
$\Omega_f$. Fig.~\ref{shortPulses} represents the pulse behavior for
 four different positions of $\Omega_f$. We
measure the degree of fillness of the pulse by the number of field oscillations
within the FWHM. One clearly sees that for filter threshold below 
the cutoff frequency, Fig.~\ref{shortPulses} a), b), the pulse is
hollow. Notice that the case of Fig.~\ref{shortPulses}b corresponds
to the cutoff frequency predicted by the "oscillating mirror" model.
Only for filter threshold positions above the spectrum cutoff given
by (\ref{ap:cutoff}) the pulse becomes filled, Fig.~\ref{shortPulses}c,d. 
These results confirm
once again the real position of the harmonic cutoff.


\section{Discussions}
\label{Discussions}


In this work we have shown analytically and numerically that the
relativistic $\gamma$-factor spikes are the physical cause for 
high harmonic generation at the boundary of overdense plasma.
It is important that the properties of these spikes are universal and follow
from the ultra relativistic similarity theory. The universal physics
of the relativistic $\gamma$-spikes inheres in the universality of
the high harmonic spectrum. 

The spectrum of the high harmonics contains the power law part
$I_n\propto1/n^{8/3}$, which goes till the cutoff at 
$\sqrt{8\alpha}\gamma^3_{max}$. Here $\gamma_{max}$ is the maximal
$\gamma$-factor and $\alpha$ describes the acceleration of the plasma boundary.
This result demonstrates that a naive "oscillating mirror" model is
insufficient for correct treatment of high harmonic generation at
plasma boundaries.

It is interesting to note though that if the plasma boundary moves
without acceleration ($\alpha\to0$) our approach restores the cutoff
$4\gamma_{max}^2$ following from the "oscillating mirror" model and
leads to $I_n\propto1/n^{5/2}$, yet this limit is irrelevant to
laser-relativistic plasma interaction.

\section*{Aknowledgements}

This work has been supported in parts by DFG Transregio~18 and by DFG 
Graduierten Kolleg~1203.


\end{document}